\documentclass[aps,prb,twocolumn,groupedaddress]{revtex4-1}

\usepackage{graphicx}
\usepackage{color}
\usepackage{ulem}

\begin{document}

\title{Superconducting fluctuations in isovalently-substituted BaFe$_2$(As$_{1-x}$P$_x$)$_2$: Possible observation of multiband effects}

\author{A. Ramos-\'Alvarez}

\author{J. Mosqueira}
\email[]{j.mosqueira@usc.es}

\author{F. Vidal}

\affiliation{LBTS, Departamento de F\'isica da Materia Condensada, Universidade de Santiago de Compostela, E-15782 Santiago de Compostela, Spain}

\author{Ding Hu}
\author{Genfu Chen}
\author{Huiqian Luo}
\author{Shiliang Li}

\affiliation{Beijing National Laboratory for Condensed Matter Physics, Institute of Physics, Chinese Academy of Sciences, Beijing 100190, People's Republic of China}

\date{\today}

\begin{abstract}
The nature of superconducting fluctuation effects in the isovalently-substituted iron pnictide BaFe$_2$(As$_{1-x}$P$_x$)$_2$ ($x\approx0.35$) is probed through measurements of the magnetization and magnetoconductivity around the superconducting transition. The results, obtained with magnetic fields up to 9~T applied in the two main crystal directions, are consistent with anisotropic Ginzburg-Landau (GL) approaches for finite applied magnetic fields. The analysis allowed to determine with accuracy the out-of-plane, $\xi_c(0)$, and in-plane, $\xi_{ab}(0)$, GL coherence lengths. Significant differences are found between the $\xi_c(0)$ values resulting from electrical transport and magnetization data. According to recent theoretical approaches, these differences could be interpreted in terms of the multiband nature of this material. The analysis of data in the low field region around the transition temperature also suggests that phase fluctuations, although possibly relevant in other Fe-based superconductors, may play a negligible role in this compound.
\end{abstract}


\maketitle

\section{Introduction}

Superconductivity appears in iron pnictides when the antiferromagnetic phase is suppressed by chemical doping of the parent compound or by applying pressure.\cite{reviews} It is now well known that an isovalent chemical dilution (e.g., the partial replacement of As by P, or Fe by Ru)\cite{09Jiang, 10Shishido,10Sharma} presents notable differences with charge doping (e.g., by partially replacing O by F in \mbox{ROFeAs}, where R is a rare earth element,\cite{08Kito,08Ren,08Chen,08Takahashi,08Kamihara} or Ba by K in BaFe$_2$As$_2$)\cite{08Rotter}. 
While isovalently-diluted compounds are closer to the clean limit,\cite{Demirdis} charged dopants create strong scattering potentials that affect superconducting properties like the vortex pinning,\cite{10Cornelis} the upper critical field,\cite{Jaroszynski,Golubov,Gurevich} and even the superconducting gap symmetry.\cite{10Cornelis,Glatz,09Onari,Kontani} In fact, many studies indicate that iron-pnictides with charged dopants present a fully gapped Fermi surface (see e.g., SmFeAsO$_{1-x}$F$_x$,\cite{09Malone} PrFeAsO$_{1-y}$ \cite{09Hashimoto2}, BaFe$_{2-x}$Co$_x$As$_2$,\cite{Barannik} Ba$_{1-x}$K$_x$Fe$_2$As$_2$,\cite{Watanabe} and BaFe$_{2-x}$Ni$_x$As$_2$ \cite{lambda,Abdel}), while isovalently substituted compounds like Ba(Fe$_{1-x}$Ru$_x$)$_2$As$_2$ and BaFe$_2$(As$_{1-x}$P$_x$)$_2$ present evidences of nodes.\cite{Yamashita,HashimotoPRB,KimPRB,Suzuki,BaFeRuAs,Yoshida,Mizukami}
It is expected that fluctuation effects around the superconducting transition  may also present differences in these two types of iron pnictides.\cite{introductionSupercon} 
These effects have been already investigated in presence of charge-doping,\cite{pallecchi,salemsugui,choi,putti,liuPLA10,kim,liuSSC11,mosqueira,welpPRB11,liu2,prando,marra,BaFeNiAssigma,mosqueiraJSNM13,
salemsuguiSST13,BaFeNiAssigma2,AhmadSST14,lascialfariBossoni,BaFeNiAsanisotropy} 
but to our knowledge still remain unexplored in isovalently substituted iron-pnictides. 

The main aim of the present work is twofold: on the one hand, we will probe the nature of superconducting fluctuation effects in isovalently-diluted compounds, and the applicability of the existing Ginzburg-Landau (GL) approaches in the different regions of the $H-T$ phase diagram. On the other hand, these effects will be used to obtain precise information about superconducting parameters like the coherence lengths (or the upper critical fields), the anisotropy factors, and the system dimensionality (a technique sometimes referred to as \textit{fluctuation spectroscopy}\cite{FlucSpectroscopy}). Fluctuation effects may also provide information about the multiband nature of these materials. In fact, it has been shown that a two-band model predicts a change in the relative amplitude of different fluctuation observables in relation with the single-band case, mainly through a renormalization of the $c$-axis coherence length.\cite{14koshelev} Among the isovalently-substituted iron pnictides, BaFe$_2$(As$_{1-x}$P$_x$)$_2$ with $x\approx0.35$ presents the highest critical temperature ($T_c\sim30$~K) and thus is the best candidate to investigate superconducting fluctuations in these materials. In this work we present a systematic study of fluctuation effects around $T_c$ in the electrical conductivity and magnetization of several high-quality single crystals of this compound. The measurements are performed in magnetic fields up to 9~T, which close to $T_c$ is well inside the finite-field (or Prange) fluctuation regime.\cite{Prange} The results will allow to probe the applicability of GL approaches for finite applied magnetic fields, and to investigate possible differences in the coherence length amplitudes associated with the multiband nature of this material. Finally, the low-field ($\sim10^{-3}$~T) behavior of the magnetization for temperatures around $T_c$ will be studied to check the relevance of phase fluctuations in BaFe$_2$(As$_{1-x}$P$_x$)$_2$, which were claimed to be important in other iron pnictides.\cite{prando,lascialfariBossoni}

\begin{figure}[t]
\includegraphics[scale=.4]{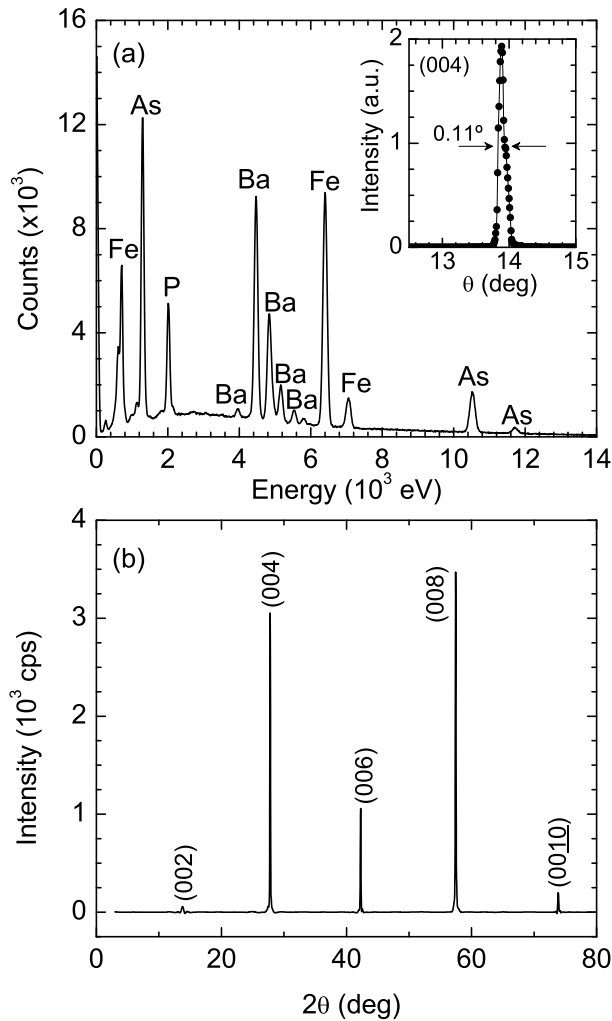}
\caption{\label{edxrx}a) Example of a typical EDX spectrum. b) Example of x-ray diffraction pattern on a single crystal, obtained by using the geometry to observe the reflections by the $ab$ layers. Inset: rocking curve associated to the $(004)$ reflection, showing that the dispersion in the orientation of the crystal $c$-axis is about 0.1$^\circ$.}
\end{figure}

\section{Growth and characterization of the crystals}

The BaFe$_2$(As$_{1-x}$P$_x$)$_2$ single crystals used in this work were grown by using the Ba$_2$As$_3$ / Ba$_2$P$_3$ self-flux method described in Ref.~\onlinecite{Nakajima15}. Some details of their characterization may be seen in Ref.~\onlinecite{Hu15}. They are plate-like, with typical surfaces of several mm$^2$ and thicknesses up to $\sim0.1$~mm. Their stoichiometry was checked with a scanning electron microscope (Zeiss FESEM Ultra Plus) equipped with a EDX (Energy Dispersive x-ray) spectroscope. A typical EDX spectrum is shown in Fig.~\ref{edxrx}(a). The average stoichiometry resulted to be Ba$_{1.04}$Fe$_{1.91}$As$_{1.33}$P$_{0.72}$, with a variation smaller than 0.4\% between the different crystals and the different studied areas. The partial substitution of As by P is about $35$~\%, which is close to the value that maximizes $T_c$.
The crystallographic structure was studied in some of the crystals by x-ray diffraction (XRD) by using a Rigaku MiniFlex II diffractometer with a Cu-target. A typical example of the reflections by the $ab$ planes is presented in Fig.~\ref{edxrx}(b). The absence of reflections other than the $(00l)$ indicates an excellent structural quality of the crystals. An example of the rocking curve for the (004) reflection is presented in the inset of Fig.~\ref{edxrx}(b). It confirms that the crystal $c$ axis presents a dispersion of only $\sim0.11^\circ$. Powder x-ray diffraction in some grounded crystals allowed to determine the lattice constants of the tetragonal structure, which resulted to be $a=b=0.39249(15)$~nm and $c=1.2833(4)$~nm, in agreement with data in the literature.\cite{Swee, 09Jiang, Ishikado}

\begin{table}[t]
\caption{\label{tablecrystals}
Physical parameters of the crystals. R1 and R2 were used in the resistivity measurements, and the stack in the magnetization measurements.}
\begin{ruledtabular}
\begin{tabular}{cccc}
crystal & surface & thickness &  mass \\ 
  & (mm$^2) $ & ($\mu$m) & (mg) \\
\hline 
R1 & $0.78\times0.49$ & 15 & 0.036 \\ 
R2 & $1.60\times1.30$ & 18 & 0.234  \\ 
Stack & $\sim$5.20 & 879 & 29.514 \\ 
\end{tabular}
\end{ruledtabular}
\end{table}

\section{Resistivity measurements}

The in-plane resistivity (along the \textit{ab} layers), $\rho_{ab}$, was measured in two single crystals with a Quantum Design's Physical Property Measurement System (PPMS), by using four contacts with an in-line configuration and an excitation current of $\sim 1$~mA at 23 Hz. The measurements were performed with magnetic fields up to $\mu_0H=9$~T applied both parallel and perpendicular to the $ab$ layers. The size of the crystals chosen for these measurements are presented in Table~\ref{tablecrystals}. The finite size of the electrical contacts leads to an uncertainty in the $\rho_{ab}$ amplitude of $\sim25$\%.

\begin{figure}[t]
\begin{center}
\includegraphics[scale=.5]{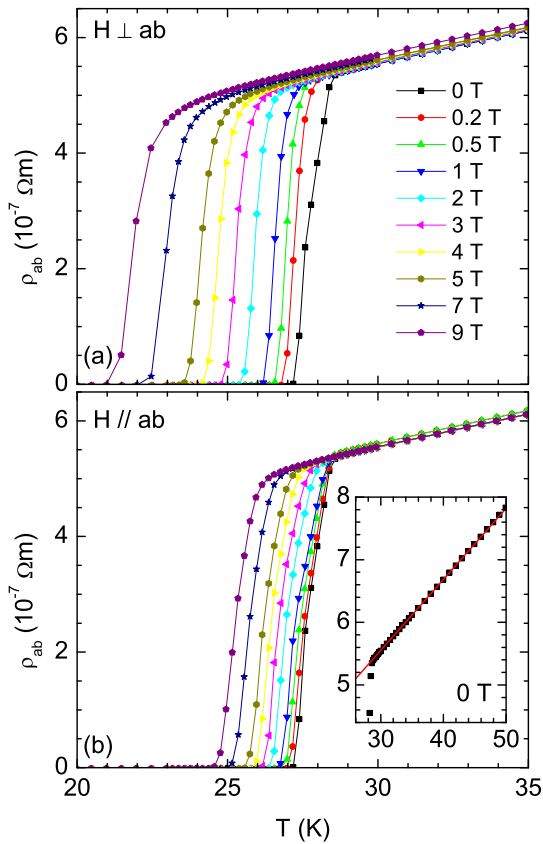}
\caption{Example (corresponding to crystal 2) of the temperature dependence of the in-plane resistivity around $T_c$, for magnetic fields up to 9~T applied in the two main crystal directions. Inset in (b): detail of the normal-state behavior for $H=0$ up to 50~K. The line is the background contribution obtained by a linear fit above 35~K. No appreciable rounding associated with fluctuations is observed above the resistive transition.}
\label{rho}
\end{center}
\end{figure}

An example (corresponding to crystal 2) of the $\rho_{ab}(T)_H$ behavior around the superconducting transition is presented in Fig.~\ref{rho}. In the overview shown in the inset it may be seen a linear temperature dependence up to 100~K, without the kinks associated to structural and magnetic transitions typical of underdoped samples.\cite{Nakajima15} In absence of an applied field, the resistive transition in both crystals is very sharp: the transition midpoint is about $\sim28$~K with a full width of 0.8~K as estimated from the 90\%-10\% criterion. This would allow to investigate fluctuation effects in a wide temperature range above the transition. However, the resistivity rounding just above $T_c$ typical of fluctuations is almost inappreciable (see the inset in Fig.~\ref{rho}(b)). This is consistent with the relatively large normal-state (or background) in-plane conductivity $\sigma_{ab}^B$, as compared with the fluctuation-induced conductivity predicted by the Aslamazov-Larkin approach for 3D materials,
\begin{equation}
\sigma_{ab}^{\rm fl}=\frac{e^2}{32\hbar\xi_c(0)}\varepsilon^{-1/2},
\label{AL}
\end{equation} 
where $\varepsilon=\ln(T/T_c)$ is the reduced temperature, $e$ is the electron charge, $\hbar$ the reduced Planck constant, and $\xi_c(0)$ the $c$-axis coherence length amplitude. By using the $\xi_c(0)=1$~nm (see below), at $\varepsilon=0.1$ one obtains $\sigma_{ab}^{\rm fl}\approx2\times10^4\;(\Omega$m)$^{-1}$, which is $10^2$ times smaller than $\sigma_{ab}^B\approx1.8\times10^6\;(\Omega$m)$^{-1}$ just above $T_c$. It is worth noting that under the largest $H$ amplitudes a slight rounding may still be observed in the upper part of the $\rho_{ab}(T)$ curves, that may be attributed to critical fluctuations near the $H_{c2}(T)$ line.

\begin{table*}[t]
\caption{\label{tableparameters}Superconducting parameters of the crystals studied, indicating the observable used to obtain them. Note the difference between the $\xi_c(0)$ values determined from the $H$-dependence of the resistive transition, and from the analysis of fluctuation effects in the magnetization. See the main text for details.}
\begin{ruledtabular}
\begin{tabular}{cccccccc}
Crystal & Observable & $\mu_0dH_{c2}^\perp/dT$ &  $\mu_0dH_{c2}^\parallel/dT$ & $T_c$ & $\xi_{ab}(0)$ & $\xi_c(0)$ & $\gamma$ \\ 
&  & (T/K) & (T/K)& (K) & (nm) & (nm) &      \\
\hline  
R1 & $\rho(T)_H$ & -1.72(2) & -3.84(6) & 28.5(1) & 2.59(2) & 1.16(4) & 2.23(6) \\  
R2 & $\rho(T)_H$ & -1.72(2) & -4.05(9) & 27.3(2) & 2.65(3) & 1.13(5) & 2.35(8) \\ 
\hline
Stack & $M_{\rm fl}(T)_H$ (Gaussian region) & -1.81(2) & -3.16(12) & 28.1 & 2.54(3) & 1.45(6) & 1.75(5) \\  
Stack & $M_{\rm fl}(T)_H$ (critical region) & -1.75(15) & -3.2(3) & 28.2(1) & 2.6(1) & 1.4(3) & 1.8(3) \\ 
\end{tabular}
\end{ruledtabular}
\end{table*}

\begin{figure}[t]
\begin{center}
\includegraphics[scale=.5]{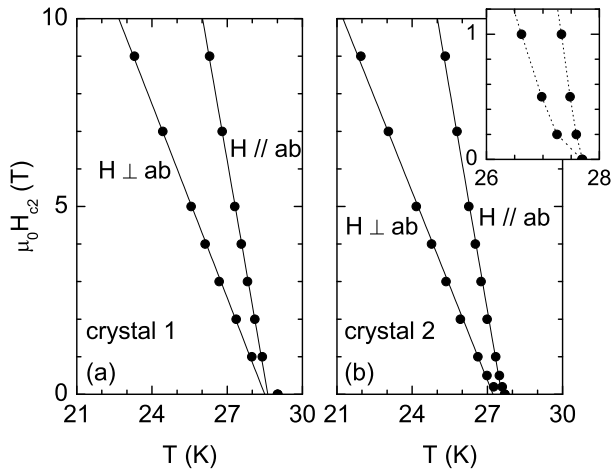}
\caption{Temperature dependence of the upper critical field for $H\perp ab$ and $H\parallel ab$, and for the two crystals studied. These data were obtained from the midpoint of the corresponding resistive transitions. Inset in (b): detail of the non linear behavior below 1~T.}
\label{Hc2}
\end{center}
\end{figure}

The absence of important fluctuation effects in the resistivity allows to determine with accuracy the temperature dependence of the upper critical fields from the transition midpoints. The result for the two studied samples is presented in Fig.~\ref{Hc2}. For both field orientations $H_{c2}(T)$ is linear to a very good approximation, although for fields below $\sim1$~T a slight positive curvature is observed (see the inset in Fig.~\ref{Hc2}(b)). This effect occurs at temperatures within the transition width and could be due to a $T_c$ distribution. According to Ref.~\onlinecite{14koshelev} it could also be attributed to the multiband nature of this material, if the upper critical field of the band with the largest coherence length is about $\sim 1$~T. Note, however, that in the general case the different bands cannot be directly associated with different coherence lengths.\cite{Babaev1,Babaev2} The solid lines in Fig.~\ref{Hc2} are linear fits for $H\geq 1$~T. The resulting $H_{c2}(T)$ slopes and the extrapolated $T_c$ values are summarized in Table~\ref{tableparameters}. From these values, the in-plane and out-of-plane GL coherence length amplitudes, $\xi_{ab}(0)$ and $\xi_c(0)$ respectively, were obtained through
\begin{equation}
\xi_{ab}(0)=\sqrt{\frac{\phi_0}{2\pi T_c\mu_0|dH_{c2}^\perp/dT|}}
\label{param1}
\end{equation}
and 
\begin{equation}
\xi_c(0)=\xi_{ab}(0)/\gamma,
\label{param2}
\end{equation}
where 
\begin{equation}
\gamma=\frac{dH_{c2}^\parallel/dT}{dH_{c2}^\perp/dT}
\label{param3}
\end{equation}
is the superconducting anisotropy factor. The resulting values are also compiled in Table~\ref{tableparameters}. 
They are expected to be inappreciably affected by the small $T_c$ distribution: i) The percolative nature of the resistivity measurements would increase the measured $T_c$ value to be introduced in Eq.~(\ref{param1}). But this effect is expected to be $\sim0.4$~K (the transition half-width, see above), less than 2\%. ii) The magnetic field displaces the $\rho(T)_H$ curves to lower temperatures without introducing an appreciably broadening (apart from the rounding observed in the upper part of the curves for $H\parallel c$). Then, the $H_{c2}(T)$ slopes determined from the transition midpoints (to be introduced in Eqs.~(\ref{param1}) and (\ref{param3})) are expected to be unaffected by a $T_c$ distribution. The linear $H_{c2}(T)$ behavior obtained above 1~T for both field orientations confirms this point.

\begin{figure}[t]
\begin{center}
\includegraphics[scale=.5]{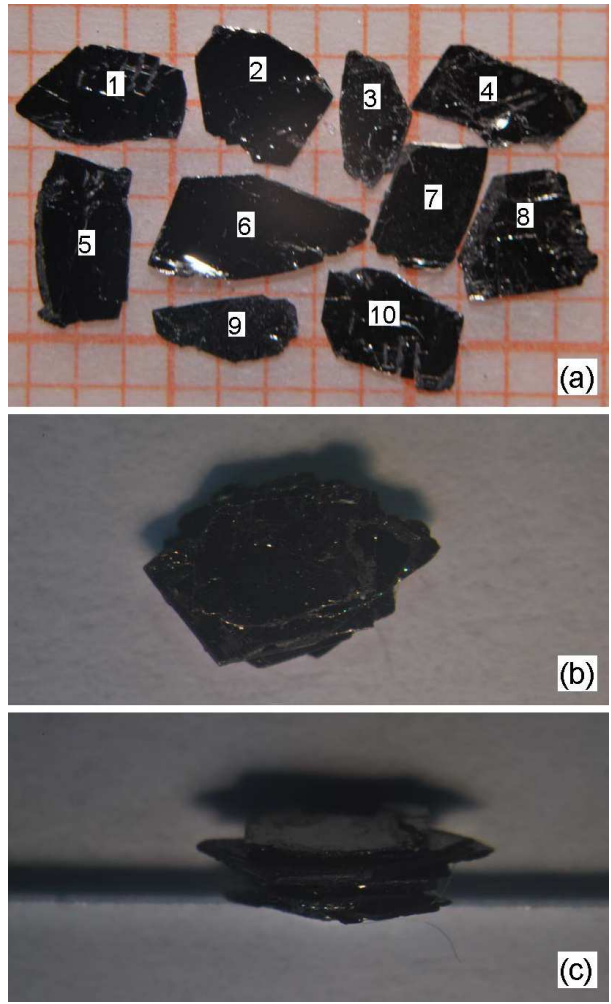}
\caption{a) Single crystals used in the magnetization measurements. b) and c) Top view and, respectively, side view of the pile formed with the single crystals shown in (a).}
\label{crystals}
\end{center}
\end{figure}

\section{Magnetization measurements}

The magnetization measurements were performed with a Quantum Design's SQUID magnetometer (model MPMS-XL) in a stack of ten high-quality single crystals glued with a minute amount of GE varnish, with a total mass of 29.514 mg and a volume of 4.71~mm$^3$ as determined from their theoretical density. A picture of the individual crystals used in the experiments and of the resulting pile may be seen in Fig.~\ref{crystals}. As we will see below, such a large sample is necessary to attain the resolution needed to study the weak diamagnetism due to superconducting fluctuations. 
By using optical microscopy we estimated that the misalignment between the different crystals in the stack is less than $2^\circ$. As the crystals anisotropy is moderate ($\gamma\approx1.8$, see Table~\ref{tableparameters}) the effect of such misalignment is negligible: the relevant parameter is the upper critical field, which according to the GL theory depends on the orientation (relative to the $c$ axis) as $H_{c2}(\theta)=H_{c2}^\perp/\sqrt{\cos^2\theta+\gamma^{-2}\sin^2\theta}$. Then, the error in the $H_{c2}$ value is as small as 0.04\% (for $H\parallel c$) and 0.14\% (for $H\parallel ab$). Even in the case that the misalignment were as large as $10^\circ$, the errors would be only 1\% (for $H\parallel c$) and 3\% (for $H\parallel ab$).

\begin{figure}[t]
\begin{center}
\includegraphics[scale=.5]{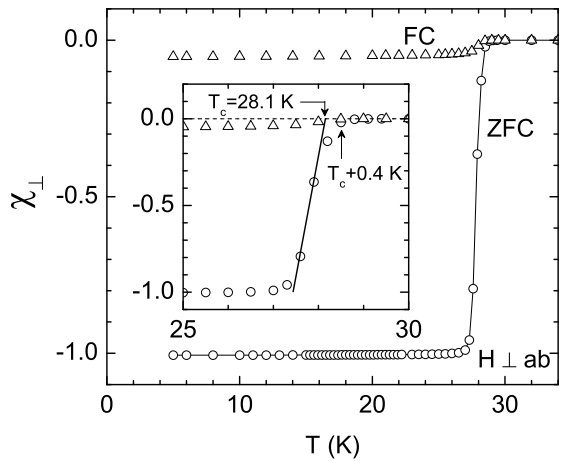}
\caption{Temperature dependence of the low field (0.2 mT) magnetic susceptibility of the pile of single crystals. These measurements were performed with $H\perp ab$ after zero-field cooling (ZFC, circles) and field-cooling (FC, triangles). The data are already corrected for demagnetizing effects. Inset: detail around $T_c$ showing the sharp diamagnetic transition (only 0.8~K full width). See the main text for details.}
\label{Tc}
\end{center}
\end{figure}

The temperature dependence of the zero-field-cooled (ZFC) magnetic susceptibility, measured with a low field ($0.2$~mT) perpendicular to the $ab$ layers, is presented in Fig.~\ref{Tc}. These data are corrected for demagnetizing effects by using the demagnetizing factor needed to attain the ideal value of -1 at low temperatures ($D=0.71$), which is consistent with the physical dimensions of the pile. From these curves, $T_c$ was estimated by a linear extrapolation to $\chi=0$ of the higher-slope $\chi(T)$ data, and the transition half-width as $\Delta T_c=T_{c0}-T_c$, where $T_{c0}$ is the highest temperature at which a diamagnetic signal is resolved (the procedure is detailed in the inset in Fig.~\ref{Tc}). In spite of the large volume of the sample, the $\Delta T_c$ value is only 0.4~K (the full width being 0.8~K), confirming its excellent stoichiometry, and allowing to study fluctuation effects in a wide temperature region above $T_c(H)$. 

\begin{figure*}[t]
\begin{center}
\includegraphics[scale=.6]{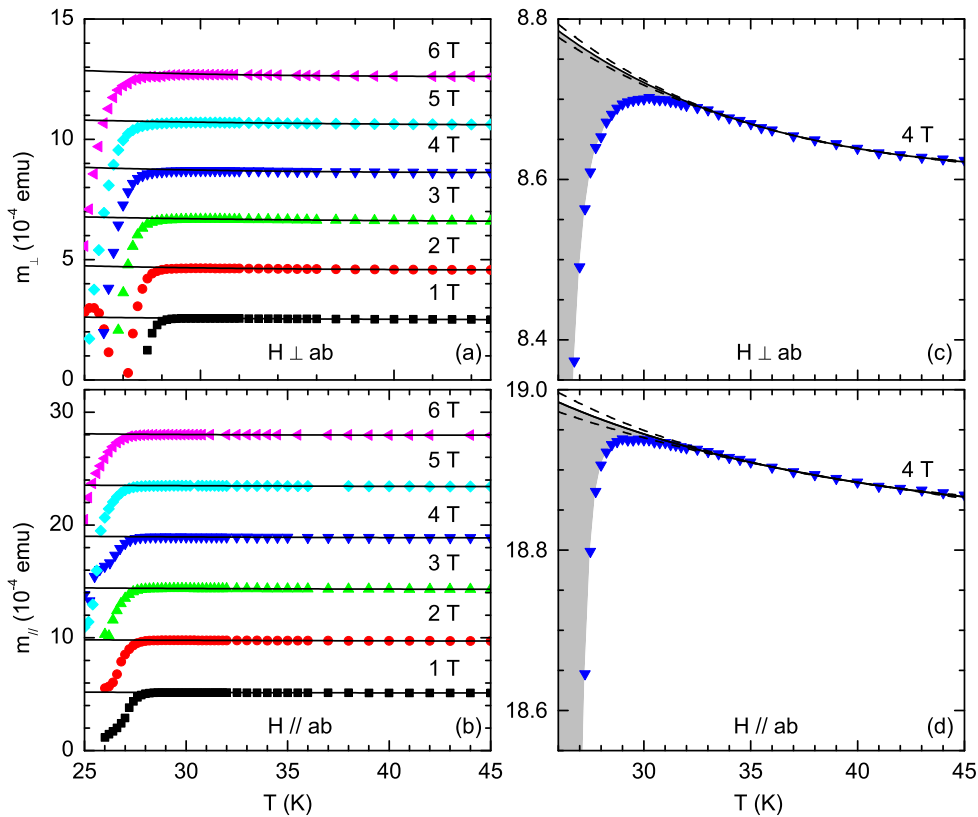}
\caption{a,b) Overview of the temperature dependence of the magnetic moment above $T_c$ for all magnetic field amplitudes and orientations investigated. The solid lines are the background contributions obtained by fitting a Curie-like function above 35~K. c,d) Detail around $T_c$ corresponding to the measurement with $\mu_0H=4$~T. The shaded areas represent the fluctuation effects, extending up to 5~K above $T_c$ and with an amplitude clearly anisotropic. The dashed lines indicate the uncertainty in the background contribution.}
\label{back}
\end{center}
\end{figure*}

To measure the weak magnetic moment due to superconducting fluctuations above $T_c$ ($m\sim -10^{-5}$ emu, see below) we used the \textit{Reciprocating Sample Option} (RSO). We averaged eight measurements consisting of 10 cycles at 1 Hz frequency, which lead to a resolution in the $\sim10^{-8}$ emu range. The magnetic fields used in the experiments range from 1 to 6 T which allowed to deeply penetrate in the finite-field (or Prange) fluctuation regime (see below). The $m(T)$ data around $T_c$ for all fields amplitudes and orientations studied are presented in Figs.~\ref{back}(a) and (b) (already subtracted from the signal of an epoxy piece used to position the sample). In the detail of Figs.~\ref{back}(c) and (d), corresponding to an applied field of 4~T, it may be appreciated a rounding extending from $T_c\approx28.1$~K to $\sim$33~K. This interval is well beyond the transition half-width ($\sim$0.4~K), so the rounding may be attributed to superconducting fluctuations. This effect prevents determining the superconducting parameters as in the case of the resistivity, and fluctuation effects will be taken into account to analyze the magnetization data.

The fluctuation contribution to the magnetic moment was determined through
\begin{equation}
m_{\rm fl}(T)=m(T)-m_B(T),
\end{equation}
where $m_B(T)$ is the background contribution coming from the samples normal state and to some extent from the sample holder. It was obtained by fitting a Curie-like function
\begin{equation}
m_B(T)=A+BT+\frac{C}{T}
\label{eqback}
\end{equation}
to the raw data in a region between 35~K up to 42~K ($A$, $B$ and $C$ are free parameters). The upper limit was chosen to avoid a small upturn (in the scale of $10^{-7}$~emu) observed above $\sim$45~K in all the measurements, and that may be attributed to an unavoidable amount of oxygen in the sample space. This upturn cannot be removed by successively pumping and venting with helium the sample space. The resulting fitting parameters followed a linear $H$-dependence within their uncertainty.

The resulting $m_B(T)$ contributions are presented as solid lines in Fig.~\ref{back}. The dashed lines represent the uncertainty associated to the fitting procedure. The resulting fluctuation magnetic susceptibility, $\chi_{\rm fl}(T)=m_{\rm fl}(T)/HV$ (where $V$ is the sample volume), is presented in Fig.~\ref{fluc}. 
The uncertainties associated to the background are only $\sim1$\% (for $H\parallel c$) and $\sim5$\% (for $H\parallel ab$) in the lower bound of the Gaussian region (solid data points in Fig.~\ref{fluc}), and negligible in the critical region. Although, as expected, they are significant for temperatures near $T_{\rm onset}$: $\sim30$\% for $H\parallel c$, and $\sim100$\% for $H\parallel ab$ at 30~K.  
Noticeably, $\chi_{\rm fl}$ is anisotropic, being significantly larger in amplitude when $H\perp ab$. Also, the $\chi_{\rm fl}$ amplitude decreases with $H$, which is an indicative that the fields used in the experiments are large enough as to enter in the finite field (or Prange) fluctuation regime.\cite{soto04} The quantitative analysis of the data would then require using theoretical approaches valid beyond the zero-field (or Schmidt) limit. 

\section{Analysis of fluctuation effects in the magnetization}

\subsection{Gaussian region above $T_c(H)$}

It has been shown that the presence of several bands contributing to the superconductivity in these materials may affect the fluctuation-induced observables through a renormalization of the coherence length amplitudes, but without appreciably affecting their functional form with respect to the single band case.\cite{14koshelev} Thus, our $\chi_{\rm fl}(T,H)$ data will be analyzed in terms of a GL approach for single-band three-dimensional anisotropic superconductors (3D-aGL approach). The possible presence of multiband effects will be probed through differences between the resulting coherence lengths and the ones previously determined from the field dependence of the resistive transition. 

In terms of the 3D-aGL approach the fluctuation magnetization $M_{\rm fl}$ of an anisotropic superconductor in presence of a finite applied magnetic field is given by \cite{carballeira3D,Klemm80,Blatter92,Hao92}
\begin{eqnarray}
&&M^{\perp}_{\rm fl}(T,H)=-\frac{k_BT\gamma}{\pi\phi_0\xi_{ab}(0)}\int_{0}^{\sqrt{c-\varepsilon}}dq\left[\frac{c-\varepsilon}{2h}\right.\nonumber\\
&-&\ln\Gamma\left(\frac{\varepsilon+h+q^2}{2h}\right)+\left(\frac{\varepsilon+q^2}{2h}\right)\psi\left(\frac{\varepsilon+h+q^2}{2h}\right)\nonumber\\
&+&\left.\ln\Gamma\left(\frac{c+h+q^2}{2h}\right)-\left(\frac{c+q^2}{2h}\right)\psi\left(\frac{c+h+q^2}{2h}\right)\right]
\label{prange}
\end{eqnarray}
for $H\perp ab$, and 
\begin{equation}
M_{\rm fl}^{\parallel}(T,H)=\frac{1}{\gamma}M_{\rm fl}^{\perp}(T,H/\gamma).
\label{prangepara}
\end{equation}
for $H\parallel ab$. Here $\Gamma$ and $\psi$ are, respectively, the gamma and digamma functions, $\varepsilon=\ln(T/T_c)$ the reduced temperature, $h=H/[\phi_0/2\pi\mu_0\xi_{ab}^2(0)]$ the reduced magnetic field, and $c$ is a total-energy cutoff constant.\cite{FVidal} The cutoff was introduced to take into account short-wavelength effects, which may be relevant in particular at high reduced magnetic fields or temperatures.\cite{introductionSupercon} In view of Eq.~(\ref{prange}), $c$ equals the reduced temperature at which fluctuation effects appear, $T_{\rm onset}\approx33$~K. So, in subsequent analyses we will use $c=\ln(T_{\rm onset}/T_c)\approx0.16$, a value slightly smaller than the one found in other Fe-based superconductors.\cite{BaFeNiAssigma,BaFeNiAssigma2,BaFeNiAsanisotropy}

\begin{figure}[t]
\begin{center}
\includegraphics[scale=.5]{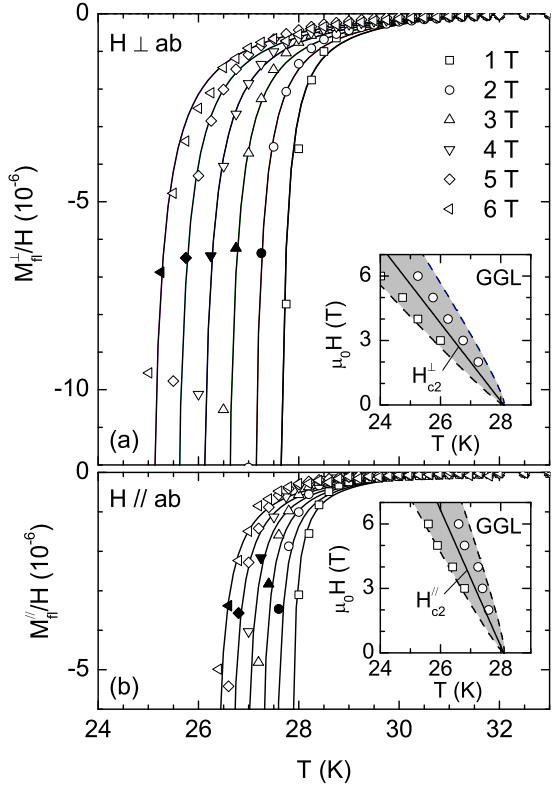}
\caption{Temperature dependence of fluctuation-induced magnetic susceptibility in the Gaussian region for both $H\perp ab$ (a) and $H\parallel ab$ (b). The lines are the best fit of Eqs.~(\ref{prange}) and (\ref{prangepara}), respectively, by using $\xi_{ab}(0)$ and $\gamma$ as free parameters. The solid data points are the lower bound of the fitting region. Insets: $H-T$ phase diagrams for both field orientations. The shaded areas represent the critical regions, as obtained from Eqs.~(\ref{criterioperp}) and (\ref{criteriopara}). The circles (squares) represent the lower bound for the applicability of the Gaussian (critical) GL approach.}
\label{fluc}
\end{center}
\end{figure}

\begin{figure}[t]
\begin{center}
\includegraphics[scale=.5]{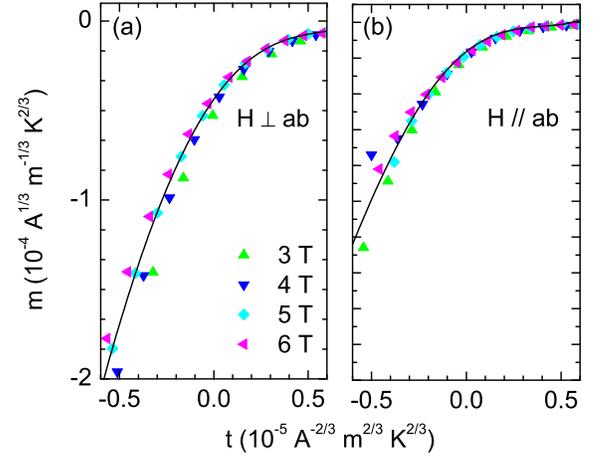}
\caption{3D-GL scaling of the magnetization in the critical region for $H\perp ab$ (a) and $H\parallel ab$ (b). The line in (a) is the experimental scaling function for $H\perp ab$, while the one in (b) for $H\parallel ab$ was calculated from the former through Eq.~(\ref{scalfunc}).}
\label{scaling}
\end{center}
\end{figure}

Eqs.~(\ref{prange}) and (\ref{prangepara}) are expected to be applicable down to the critical fluctuation region. This last is bounded by the so-called $H$-dependent Ginzburg criterion, which for 3D anisotropic superconductors may be written as,\cite{ikeda89,ikeda90,kim92}
\begin{equation}
T_G^\perp(H)\approx T_{c}^\perp(H) \pm T_c\left[\frac{4\pi k_B\mu_0H}{\Delta c\xi_c(0)\phi_0}\right]^{2/3}
\label{criterioperp}
\end{equation}
for $H\perp ab$, and  
\begin{equation}
T_G^\parallel(H)\approx T_{c}^\parallel(H)\pm T_c\left[\frac{4\pi k_B\mu_0H}{\Delta c\xi_c(0)\gamma\phi_0}\right]^{2/3}
\label{criteriopara}
\end{equation}
for $H\parallel ab$, where $\Delta c$ is the specific-heat jump at $T_c$. 
Let us finally note that Eqs.~(\ref{prange}) and (\ref{prangepara}) were already successfully used to explain the susceptibility rounding above $T_c$ of iron pnictides like Ba$_{1-x}$K$_x$Fe$_2$As$_2$ and Ba(Fe$_{1-x}$Ni$_x$)$_2$As$_2$,\cite{mosqueira,BaFeNiAsanisotropy} and of other 3D anisotropic superconductors like MgB$_2$ and NbSe$_2$.\cite{MgB2,NbSe2} In turn, the 2D and 2D$-$3D (Lawrence-Doniach) versions accounted for the behavior of different high-$T_c$ cuprates.\cite{MosqueiraEPL,Tl2223,TlPb1212,intrinsic} 

Eqs.~(\ref{prange}) and (\ref{prangepara}) were fitted to the set of $M_{\rm fl}/H$ data for both $H\perp ab$ and $H\parallel ab$ with only two free parameters, $\xi_{ab}(0)$ and $\gamma$. The best fit is represented as solid lines in Fig.~\ref{fluc}. The fitting interval range from $T_{\rm onset}$ to the solid data points. If the fitting interval is extended to lower temperatures, the fit quality is considerably worsened. In fact, these solid data points are already close to the upper bound of the critical fluctuation region. This is better seen in the $H-T$ phase diagrams in the insets of Fig.~\ref{fluc}, where the limit of applicability of the GGL approach is indicated by circles, and the shaded areas represent the critical regions evaluated with Eqs.~(\ref{criterioperp}) and (\ref{criteriopara}) by using $\Delta c/T_c\sim0.1$~J/molK$^2$,\cite{Zaanen} and the $\xi_c(0)$ and $\gamma$ values in Table~\ref{tableparameters}. 

The values for the fitting parameters, $\xi_{ab}(0)$ and $\gamma$, are compiled in Table~\ref{tableparameters}, together with the $\xi_c(0)$ value obtained as $\xi_{ab}(0)/\gamma$. The indicated uncertainties account for the range in which the fitting parameters may be changed without appreciably worsen the fitting quality. The $\xi_{ab}(0)$ value is in excellent agreement with the one determined in Section III from resistivity measurements, but the $\xi_c(0)$ value is a 30~\% larger. Such a difference is beyond the experimental uncertainty, and will be discussed in Sec.~V.C.

\subsection{Fluctuation diamagnetism in the critical region around $T_c(H)$}

As a check of consistency of the above analysis we studied the data in the critical region around the $H_{c2}(T)$ line, where the Gaussian approximation breaks down.\cite{introductionSupercon} In this region, the 3D-GL approach in the lowest-Landau-level approximation predicts that $M_{\rm fl}^\perp(T,H)$ and $M_{\rm fl}^\parallel(T,H)$ follow a scaling behavior, $m_{\perp,\parallel}=f_{\perp,\parallel}(t_{\perp,\parallel})$, the scaling variables being \cite{Ullah1,Ullah2}
\begin{equation}
m_{\perp,\parallel}\equiv \frac{M_{\perp,\parallel}}{(HT)^{2/3}}
\label{varm}
\end{equation}
and
\begin{equation}
t_{\perp,\parallel}\equiv \frac{T-T_{c}^{\perp,\parallel}(H)}{(HT)^{2/3}}.
\label{vart}
\end{equation}
As always, the indexes $\perp$ and $\parallel$ stand for $H\perp ab$ and $H\parallel ab$, respectively. The scaling functions for $H\perp ab$ and $H\parallel ab$ are related through\cite{mosqueira}
\begin{equation}
f_\parallel(t_\parallel)=\frac{f_{\perp}(t_{\parallel}\gamma^{2/3})}{\gamma^{5/3}}.
\label{scalfunc}
\end{equation}

By assuming a linear temperature dependence of the upper critical fields, the scalings for both $H\perp ab$ and $H\parallel ab$ depend on only three parameters: $T_c$, and the $H_{c2}^\perp(T)$ and $H_{c2}^\parallel(T)$ slopes. The best scalings, presented in Fig.~\ref{scaling}, were obtained with the values for these parameters presented in the last row of Table~\ref{tableparameters}. The indicated uncertainties represent the range of values for which the scalings are not appreciable worsened. The corresponding $\xi_{ab}(0)$ and $\xi_c(0)$ values, obtained by using Eqs.~(\ref{param1}-\ref{param3}), are also compiled in Table~\ref{tableparameters}. In spite of the larger uncertainties, these values are consistent with the ones found in the analysis of the Gaussian region. Other indications of the consistence of the present analysis are: i) The scaling is acceptable down to $t_{\parallel,\perp}\sim-3\times10^{-6}$, a value in good agreement with the lower bound of the critical region (it leads to the squares in the $H-T$ phase diagrams in the insets of Fig.~\ref{fluc}). ii) The experimental scaling functions (solid lines in Fig.~\ref{scaling}) are related to each other as predicted by Eq.~(\ref{scalfunc}), see the caption in Fig.~\ref{scaling}. iii) The smooth temperature dependence of the magnetization around $T_c(H)$ (i.e., around $t=0$), see Fig. 8, justifies the seemingly negligible role of the possible $T_c$-inhomogeneities in the sample.

\begin{table}[t]
\caption{\label{tableliterature}Summary of the superconducting parameters in the literature for BaFe$_2$(As$_{1-x}$P$_x$)$_2$ with $x$ near the optimal (maximum-$T_c$) value.}
\begin{ruledtabular}
\begin{tabular}{ccccccc}
$T_c$ &  $\mu_0dH_{c2}^{\perp}/dT$ & $\mu_0dH_{c2}^\parallel/dT$ & $\xi_{ab}(0)$ & $\xi_{c}(0)$& observable & Ref. \\ 
(K) & (T/K) & (T/K) & (nm)  & (nm) & &\\
\hline 
30.6 & -1.72 & -3.66 & 2.50 & 1.17 & $\rho(T)_H$ & \onlinecite{Chong}\\
30.0 &-2.1 & --&2.30&--& $C(T)_H$ & \onlinecite{HashimotoPRB}\\
29.0 &-2.1&--&2.30&--& $C(T)_H$ & \onlinecite{Putzke}\\
28.1&-2.23&-5.73&2.29&0.9& $C(T)_H$ & \onlinecite{Chaparro}\\
28.4&-2.1&--&2.33&--& $C(T)_H$ & \onlinecite{15Diaoarxiv}\\
30.5&-1.67&-2.41&2.54&1.75& LCR & \onlinecite{Swee}\\
\end{tabular}
\end{ruledtabular}
\end{table}

\subsection{Discussion of the results}

As it is shown in Table~\ref{tableparameters}, the $\xi_{ab}(0)$ values determined from the resistivity measurements are in excellent agreement with the one derived from the fluctuation magnetization. However, there is a significant difference between the corresponding $\xi_c(0)$ values. This effect cannot be attributed to an incomplete superconducting volume fraction in the stack used for the magnetization measurements, because $\xi_{ab}(0)$ and $\gamma$ (or $\xi_c(0)$) affect the amplitudes of $M_{\rm fl}^\perp$ and $M_{\rm fl}^\parallel$, but also their dependence on $H$. 
The $\xi_{ab}(0)$ and $\xi_c(0)$ values in the literature for BaFe$_2$(As$_{1-x}$P$_x$)$_2$ with $T_c$ values near the maximum one, are summarized in Table~\ref{tableliterature}. These data were obtained from the $T_c(H)$ dependence, as probed by dc and ac electrical transport and by a static property like the specific heat. The $\xi_{ab}(0)$ values in these works are close to each other and to the values obtained here (see Table~\ref{tableparameters}). The $\xi_c(0)$ value obtained from dc resistivity\cite{Chong} is also in excellent agreement with our data obtained from the same observable. However, a notable difference is found with the $\xi_c(0)$ values derived from other observables (0.9-1.75~nm). These results may be somewhat affected by the uncertainty associated to the use of a criterion to determine $T_c(H)$, but could indicate a possible dependence of $\xi_c(0)$ on the observable used to determine it. 

Our present results may be due to the presence of several bands contributing to the superconductivity in the material under study.\cite{fanfarillo,Gurevich,orlova,marciani} In fact, a recently proposed two-band model predicts that the functional forms of dominating divergences of the fluctuation-induced specific heat and electrical conductivity do not change with respect to the single-band case, but their relative amplitudes are affected mainly by a renormalization of the $c$-axis coherence length.\cite{14koshelev} It has been suggested that this effect could be behind a dependence of the upper critical fields of FeSe$_{0.5}$Te$_{0.5}$ on the observable used to determine them. In particular, the $H_{c2}$ values derived from measurements of the specific heat are systematically larger than the ones obtained from the electrical resistivity, and these last are in turn larger than the ones obtained from magnetic torque.\cite{klein,serafin,braithwaite} It would be then very interesting to check whether an expression for $M_{\rm fl}(T,H)$ in a two-band superconductor (still unavailable) would also quantitatively account for the differences found here with the data obtained from resistivity measurements.

\begin{figure}[t]
\begin{center}
\includegraphics[scale=.4]{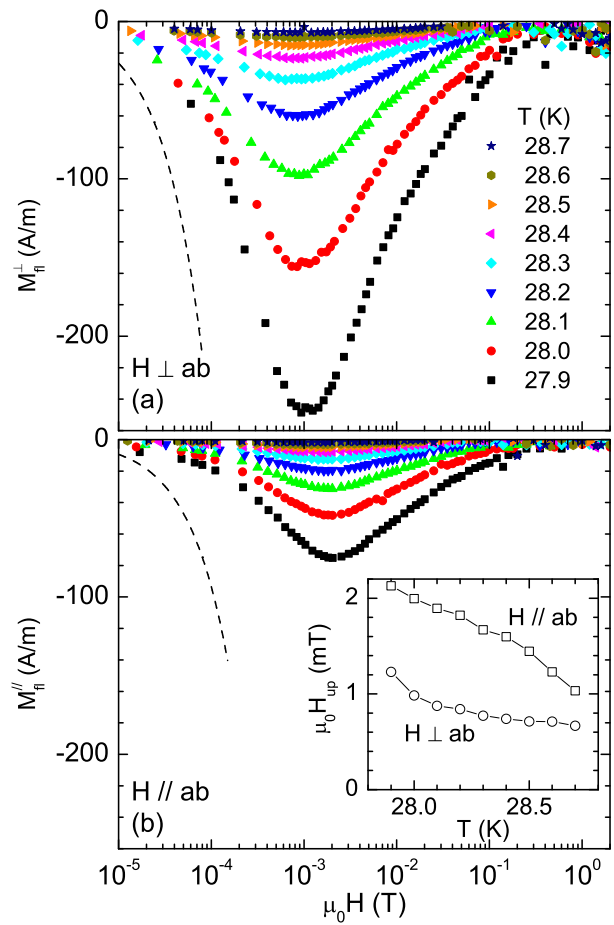}
\caption{Anomalous $H$-dependence of the fluctuation magnetization for temperatures just above $T_c$ and for both $H\perp ab$ (a) and $H\parallel ab$ (b). As a reference, the dashed lines represent the ideal diamagnetic response (evaluated taking into account the demagnetizing factors of the sample under study). The anomalous upturn appears at temperatures within the transition width (0.8~K full width). Inset: Temperature dependence of the upturn magnetic field, $H_{\rm up}$.}
\label{anomaly}
\end{center}
\end{figure}

\subsection{Enhanced diamagnetism in the low-field region just above $T_c$}

For completeness, we have also explored this region of the phase diagram. In contrast with the linear behavior predicted by Gaussian GL approaches under low fields ($h\ll \varepsilon$), the $M_{\rm fl}(H)$ isotherms just above $T_c$ present an anomalous upturn at a field $\mu_0H_{up}\approx10^{-3}$~T for both field orientations (see Fig.~\ref{anomaly}). Below $H_{\rm up}$ the magnetic susceptibility is orders of magnitude larger than the one associated to Gaussian superconducting fluctuations. Above $H_{\rm up}$ the $M_{\rm fl}$ amplitude decreases and, consistently with Fig.~\ref{fluc}, for fields about $\sim1$~T the conventional GGL behavior is recovered. 

A qualitatively similar anomalous behavior has been observed in high-$T_c$ cuprates like La$_{2-x}$Sr$_x$CuO$_4$\cite{LSCOanomaly,Lucia06,11Mosqueira} and Y$_{1-x}$Ca$_x$Ba$_2$Cu$_3$O$_y$,\cite{YCaBaCuO} and more recently in Fe-based superconductors like SmFeAsO$_{1-x}$F$_x$ (Sm1111) \cite{prando} and Ba(Fe$_{1-x}$Rh$_x$)$_2$As$_2$ (Rh122)\cite{lascialfariBossoni}. The similarities between our results and these previous works in iron pnictides are even quantitative. For instance, the isotherms measured at temperatures near $T_c$ present very similar $H_{\rm up}$ and $M_{\rm up}$ values (in the case of Sm1111 one has to take into account its 2D nature and that the sample is granular). So, to a large extent the effect in BaFe$_2$(As$_{1-x}$P$_x$)$_2$ has the same origin than in Rh122 and Sm1111. In Refs.~\onlinecite{prando} and \onlinecite{Rigamonti} it has been proposed that, like in the HTSC, the anomaly is due to the presence of important phase fluctuations. Alternatively, it has been proposed that it is a consequence of a $T_c$ distribution, which in cuprate or Fe-based superconductors would be associated to their non-stoichiometric nature.\cite{11Mosqueira} In the case of BaFe$_2$(As$_{1-x}$P$_x$)$_2$ some indications support this last explanation: On the one side, below $H_{\rm up}$ the magnetic susceptibility is a significant fraction of the perfect diamagnetism (indicated as a dashed line in Fig.~\ref{anomaly}), as if a significant volume fraction of the sample were actually in the Meissner region. On the other, the effect is limited to a range of temperatures above $T_c$ of the order of the transition width. Finally, the $H_{\rm up}$ value at the average $T_c$ is consistent with the lower critical field of the highest-$T_c$ domains, which may be approximated by $-(\partial H_{c1}/\partial T)_{T_c}\Delta T_c$, and that is in the $10^{-3}$~T range.\cite{Rutgers}

An argument supporting the possible presence of phase fluctuations in Sm1111 and Rh122 is that $H_{\rm up}$ increases with the temperature above $T_c$, while the lower critical field decreases.\cite{Bernardi,prando,lascialfariBossoni} Such an $H_{\rm up}$ increase with $T$ is not observed in BaFe$_2$(As$_{1-x}$P$_x$)$_2$ (see inset in Fig.~\ref{anomaly}), in spite that the transition width of our sample is comparable or even smaller (in relative terms) than those in Refs.~\onlinecite{prando} and \onlinecite{lascialfariBossoni}. As a consequence, we may assume that phase fluctuations do not play a role in this compound, or at least they are much less relevant than in Sm1111 and Rh122.

\section{Conclusions}

We have presented measurements of the magnetization and electrical resistivity around the superconducting transition of the isovalently-substituted iron pnictide BaFe$_2$(As$_{1-x}$P$_x$)$_2$ with $x\approx0.35$, near the optimal (maximum-$T_c$) value. The measurements were performed with magnetic fields up to 9~T in the conductivity and 6~T in the magnetization, in both cases applied in the two main crystal directions (parallel and perpendicular to the $ab$ layers). 
Due to the relatively large normal-state electrical conductivity near $T_c$, the fluctuation effects in this observable are almost unobservable. As a consequence, the resistive transitions are very well defined, allowing to determine with accuracy the temperature dependence of the parallel and perpendicular upper critical fields, and the in-plane $\xi_{ab}(0)$ and out-of-plane $\xi_c(0)$ coherence lengths. 
On the contrary, the magnetization presents an appreciable diamagnetic contribution just above $T_c$ that may be attributed to superconducting fluctuations. These data were consistently analyzed in the different fluctuation regimes (Gaussian and critical) in terms of existing Ginzburg-Landau approaches for single-band superconductors. While the resulting $\xi_{ab}(0)$ is in excellent agreement with the one obtained from the field dependence of the resistive transition, $\xi_c(0)$ is significantly larger ($\sim30$\%). 
A dependence of the coherence lengths with the observable used to determine them was also found in other Fe-based superconductors.\cite{klein,serafin,braithwaite}
Recent theoretical approaches show that the presence of several bands contributing to the superconductivity affects the fluctuation effects mainly through a renormalization of the $c$-axis coherence length.\cite{14koshelev} Thus, we suggest that our present results could be a consequence of the multiband nature of the compound under study. An expression for the fluctuation magnetization in two-band superconductors would be desirable to confirm this proposal. 
Finally, from the analysis of the anomalous upturn in the $M(H)$ isotherms just above $T_c$, we find that phase fluctuations may play a negligible role in this compound, contrary to what is proposed for other Fe-based superconductors.\cite{prando,lascialfariBossoni}

\begin{acknowledgments}

This work was supported by the Xunta de Galicia (grant no. GPC2014/038), and COST Action MP1201 (NanoSC). AR-A acknowledge financial support from Spain's MICINN through a FPI grant (no. BES-2011-046820). The work at IOP, CAS in China is supported by NSFC Program (No. 11374011), MOST of China (973 project: 2011CBA00110) and CAS (SPRP-B: XDB07020300).

\end{acknowledgments}


\begin{references}

\bibitem{reviews}For reviews see D. C. Johnston, Adv. Phys. \textbf{59}, 803 (2010); J. Paglione and R. Greene, Nature Phys. \textbf{6}, 645 (2010). 

\bibitem{09Jiang}S. Jiang, H. Xing, G. Xuan, C. Wang, Z. Ren, C. Feng, J. Dai, Z. Xu, and G. Cao, J. Phys.: Condens. Matter \textbf{21}, 382203 (2009).

\bibitem{10Shishido}H. Shishido, A. F. Bangura, A. I. Coldea, S. Tonegawa, K. Hashimoto, S. Kasahara, P. M. C. Rourke, H. Ikeda, T. Terashima, R. Settai, Y. \=Onuki, D. Vignolles, C. Proust, B. Vignolle, A. McCollam, Y. Matsuda, T. Shibauchi, and A. Carrington, Phys. Rev. Lett. \textbf{104}, 057008 (2010).

\bibitem{10Sharma}S. Sharma, A. Bharathi, S. Chandra, V. R. Reddy, S. Paulraj, A. T. Satya, V. S. Sastry, A. Gupta, and C. S. Sundar, Phys. Rev. B \textbf{81}, 174512 (2010).

\bibitem{08Kito}H. Kito, H. Eisaki, and A. Iyo, J. Phys. Soc. Jpn. \textbf{77}, 063707 (2008).

\bibitem{08Ren}Z.A. Ren, J. Yang, W. Lu, W. Yi, G. Che, X. Dong, L. Sun, and Z. Zhao, Mater. Res. Innovations \textbf{12}, 105 (2008).

\bibitem{08Chen}G. F. Chen, Z. Li, D. Wu, G. Li, W. Z. Hu, J. Dong, P. Zheng, J. L. Luo, and N. L. Wang, Phys. Rev. Lett. \textbf{100}, 247002 (2008).

\bibitem{08Takahashi}H. Takahashi, K. Igawa, K. Arii, Y. Kamihara, M. Hirano, and H. Hosono, Nature \textbf{453}, 376 (2008).

\bibitem{08Kamihara}Y. Kamihara, T. Watanabe, M. Hirano, and H. Hosono, J. Am. Chem. Soc. \textbf{130}, 3296 (2008). 

\bibitem{08Rotter}M. Rotter, M. Tegel, and D. Johrendt, Phys. Rev. Lett. \textbf{101}, 107006 (2008). 

\bibitem{Demirdis}S. Demirdi\c{s}, Y. Fasano, S. Kasahara, T. Terashima, T. Shibauchi, Y. Matsuda, M. Konczykowski, H. Pastoriza, and C. J. van der Beek, Phys. Rev. B \textbf{87}, 094506 (2013). 

\bibitem{10Cornelis}C. J. van der Beek, M. Konczykowski, S. Kasahara, T. Terashima, R. Okazaki, T. Shibauchi, and Y. Matsuda, Phys. Rev. Lett. \textbf{105}, 267002 (2010). 

\bibitem{Jaroszynski}J. Jaroszynski, F. Hunte, L. Balicas, Y.-j. Jo, I. Rai\u{c}evi\'c, A. Gurevich, D. C. Larbalestier, F. F.  Balakirev, L. Fang, P. Cheng, Y. Jia, and H. H. Wen, Phys. Rev. B \textbf{78}, 174523 (2008). 

\bibitem{Golubov}A. A. Golubov and A. E. Koshelev, Phys. Rev. B \textbf{68}, 104503 (2003). 

\bibitem{Gurevich}A. Gurevich, Phys. Rev. B \textbf{82}, 184504 (2010). 

\bibitem{Glatz}A. Glatz and A. E. Koshelev, Phys. Rev. B \textbf{82}, 012507 (2010). 

\bibitem{09Onari}S. Onari and H. Kontani, Phys. Rev. Lett. \textbf{103}, 177001 (2009). 

\bibitem{Kontani}H. Kontani and S. Onari, Phys. Rev. Lett. \textbf{104}, 157001 (2010). 

\bibitem{09Malone}L. Malone, J. D. Fletcher, A. Serafin, A. Carrington, N. D. Zhigadlo, Z. Bukowski, S. Katrych, J. Karpinski Phys. Rev. B \textbf{79}, 140501 (2009). 

\bibitem{09Hashimoto2}K. Hashimoto, T. Shibauchi, T. Kato, K. Ikada, R. Okazaki, H. Shishido, M. Ishikado, H. Kito, A. Iyo, H. Eisaki, S. Shamoto, and Y. Matsuda, Phys. Rev. Lett. \textbf{102}, 017002 (2009). 

\bibitem{Barannik}A. Barannik, N. T. Cherpak, M. A. Tanatar, S. Vitusevich, V. Skresanov, P. C. Canfield, and R. Prozorov, Phys. Rev. B \textbf{87}, 014506 (2013). 

\bibitem{Watanabe}D. Watanabe, T. Yamashita, Y. Kawamoto, S. Kurata, Y. Mizukami, T. Ohta, S. Kasahara, M. Yamashita, T. Saito, H. Fukazawa, Y. Kohori, S. Ishida, K. Kihou, C.H. Lee, A. Iyo, H. Eisaki, A.B. Vorontsov, T. Shibauchi and Y. Matsuda, Phys. Rev. B \textbf{89}, 115112 (2014).

\bibitem{lambda}R. I. Rey, A. Ramos-\'Alvarez, J. Mosqueira, S. Salem-Sugui Jr., A. D. Alvarenga, H.-Q. Luo, X.-Y. Lu, R. Zhang, and F. Vidal, Supercond. Sci. Technol. \textbf{27}, 055015 (2014).

\bibitem{Abdel}M. Abdel-Hafiez, Y. Zhang, Z. He, J. Zhao, C. Bergmann, C. Krellner, C.-G. Duan, X. Lu, H. Luo, P. Dai, and X.-J. Chen, Phys. Rev. B \textbf{91}, 024510 (2015).

\bibitem{Yamashita}M. Yamashita, N. Nakata, Y. Senshu, S. Tonegawa, K. Ikada, K. Hashimoto, H. Sugawara, T. Shibauchi, and Y. Matsuda, Phys. Rev. B \textbf{80}, 220509 (2009). 

\bibitem{HashimotoPRB}K. Hashimoto, M. Yamashita, S. Kasahara, Y. Senshu, N. Nakata,  S. Tonegawa, K. Ikada, A. Serafin, A.  Carrington, T. Terashima, H.  Ikeda, T. Shibauchi, and Y.Matsuda, Phys. Rev. B \textbf{81}, 220501 (2010).

\bibitem{KimPRB}J. S. Kim, P. J. Hirschfeld, G. R. Stewart, S. Kasahara,   T. Shibauchi, T. Terashima, and Y. Matsuda, Phys. Rev. B \textbf{81}, 214507 (2010).

\bibitem{Suzuki} K. Suzuki, H. Usui, and K. Kuroki, J. Phys. Soc. Jpn. \textbf{80}, 013710 (2011).

\bibitem{BaFeRuAs}X. Qiu, S. Y. Zhou, H. Zhang, B. Y. Pan, X. C. Hong, Y. F. Dai, M. J. Eom, J. S. Kim, Z. R. Ye, Y. Zhang, D. L. Feng, and S. Y. Li, Phys. Rev. X \textbf{2}, 011010 (2012).

\bibitem{Yoshida}T. Yoshida, S. Ideta, T. Shimojima, W. Malaeb, K. Shinada, H. Suzuki, I. Nishi, A. Fujimori, K. Ishizaka, S. Shin, Y. Nakashima, H. Anzai, M. Arita, A. Ino, H. Namatame, M. Taniguchi,H. Kumigashira, K. Ono, S. Kasahara, T. Shibauchi, T. Terashima, Y. Matsuda, M. Nakajima, S. Uchida, Y. Tomioka, T. Ito, K. Kihou, C. H. Lee, A. Iyo, H. Eisaki, H. Ikeda, R. Arita, T. Saito, S. Onari and H. Kontani, Sci. Rep. \textbf{4}, 7292 (2014).

\bibitem{Mizukami}Y. Mizukami, M. Konczykowski, Y. Kawamoto, S. Kurata, S. Kasahara, K. Hashimoto, V. Mishra, A. Kreisel, Y. Wang, P. J. Hirschfeld, Y. Matsuda and T. Shibauchi,  Nat. Commun. \textbf{5}, 5657 (2014). 

\bibitem{introductionSupercon}For reviews on the theoretical and experimental aspects of the superconducting fluctuations see M. Tinkham, \textit{Introduction to Superconductivity} (McGraw-Hill, New York, 1996), Chap. 8; F. Vidal and M. V. Ramallo, \textit{The Gap Symmentry and Fluctuations in High-$T_c$ Superconductors} (Plenum, London, 1998), p. 443; A. Larkin and A. Varlamov, \textit{Theory of fluctuations in superconductors} (Oxford University Press, Oxford, 2005).

\bibitem{pallecchi}I. Pallecchi, C. Fanciulli, M. Tropeano, A. Palenzona, M. Ferretti, A. Malagoli, A. Martinelli, I. Sheikin, M. Putti, and C. Ferdeghini, Phys. Rev. B \textbf{79}, 104515 (2009).

\bibitem{salemsugui}S. Salem-Sugui Jr., L. Ghivelder, A. D. Alvarenga, J. L. Pimentel, H. Luo, Z. Wang, and H.-H. Wen, Phys. Rev. B \textbf{80},014518 (2009).

\bibitem{choi}C. Choi, S. H. Kim, K.-Y. Choi, M.-H. Jung, S.-I. Lee, X. F. Wang, X. H. Chen, and X. L. Wang, Supercond. Sci. Technol. \textbf{22}, 105016 (2009).

\bibitem{putti}M. Putti, I. Pallecchi, E. Bellingeri, M. R. Cimberle, M. Tropeano, C. Ferdeghini, A. Palenzona, C. Tarantini, A.
Yamamoto, J.Jiang, J. Jaroszynski, F. Kametani, D. Abraimov, A. Polyanskii, J. D. Weiss, E. E. Hellstrom, A.
Gurevich, D. C. Larbalestier, R. Jin, B. C. Sales, A. S. Sefat, M. A. McGuire, D. Mandrus, P. Cheng, Y. Jia, H. H. Wen, S.
Lee, and C. B. Eom, Supercond. Sci. Technol. \textbf{23}, 034003 (2010).

\bibitem{liuPLA10}S. L. Liu, W. Haiyun, and B. Gang, Phys. Lett. A \textbf{374}, 3529 (2010).

\bibitem{kim}S. H. Kim, C. H. Choi, M.-H. Jung, J.-B. Yoon, Y.-H. Jo, X. F. Wang, X. H. Chen, X. L. Wang, S.-I. Lee,and K.-Y. Choi, J. Appl. Phys. \textbf{108}, 063916 (2010).

\bibitem{liuSSC11}S. L. Liu, W. Haiyun, and B. Gang, Solid State Comm. \textbf{151}, 1 (2011).

\bibitem{mosqueira}J. Mosqueira, J. D. Dancausa, F. Vidal, S. Salem-Sugui Jr., A. D.  Alvarenga, H.-Q. Luo, Z.-S. Wang, and H.-H. Wen, Phys. Rev. B \textbf{83}, 094519 (2011).

\bibitem{welpPRB11}U. Welp, C. Chaparro, A. E. Koshelev, W. K.  Kwok, A. Rydh, N. D.  Zhigadlo, J. Karpinski, and S. Weyeneth, Phys. Rev. B \textbf{83}, 100513 (2011).

\bibitem{liu2}S. L. Liu, G. Longyan, B. Gang, W. Haiyun, and L. Yongtao, Supercond. Sci. Technol. \textbf{24}, 075005 (2011).

\bibitem{prando} G. Prando, A. Lascialfari, A. Rigamonti, L. Roman\'o, S. Sanna, M. Putti, and M. Tropeano, Phys. Rev. B \textbf{84}, 064507 (2011).

\bibitem{marra}P. Marra, A. Nigro, Z. Li, G.F. Chen, N.L. Wang, J.L. Luo, and C. Noce, New J. Phys. \textbf{14}, 043001 (2012).

\bibitem{BaFeNiAssigma}R. I. Rey, C. Carballeira, J. Mosqueira, S. Salem-Sugui Jr., A. D. Alvarenga, H.-Q. Luo, X.-Y. Lu, Y.-C. Chen, and F. Vidal, Supercond. Sci. Technol. \textbf{26}, 055004 (2013).

\bibitem{mosqueiraJSNM13}J. Mosqueira, J. D. Dancausa, C. Carballeira, S. Salem-Sugui Jr., A. D. Alvarenga, H.-Q. Luo, Z.-S. Wang, H.-H. Wen, and F. Vidal, J. Supercond. Nov. Magn. \textbf{26}, 1217 (2013).

\bibitem{salemsuguiSST13}S. Salem-Sugui, A. D. Alvarenga, R. I. Rey, J. Mosqueira, H.-Q. Luo, and X.-Y. Lu, Supercond. Sci. Technol. \textbf{26}, 125019 (2013).

\bibitem{BaFeNiAssigma2}R. I. Rey, A. Ramos-\'Alvarez, C. Carballeira, J. Mosqueira, F. Vidal, S. Salem-Sugui Jr., A. D. Alvarenga, R.
Zhang, and H. Luo, Supercond. Sci. Technol. \textbf{27}, 075001 (2014).

\bibitem{AhmadSST14}D. Ahmad, B. H. Min, W. J. Choi, S. Salem-Sugui Jr.,  J. Mosqueira, and Y. S. Kwon, Supercond. Sci. Technol. \textbf{27}, 125006 (2014).

\bibitem{lascialfariBossoni}L. Bossoni, L. Roman\'o, P. C. Canfield, and A. Lascialfari, J. Phys.: Condens. Matter \textbf{26}, 405703 (2014).

\bibitem{BaFeNiAsanisotropy}A. Ramos-\'Alvarez, J. Mosqueira, F. Vidal, X. Lu, and H. Luo, Supercond. Sci. Technol. \textbf{28}, 075004 (2015).

\bibitem{FlucSpectroscopy}F. Vidal, C. Torr\'on, J. Vi\~na, and J. Mosqueira, Physica C \textbf{332}, 166 (2000); See also A. Glatz, A. A. Varlamov, and V. M. Vinokur, Phys. Rev. B \textbf{84}, 104510 (2011).

\bibitem{14koshelev}A. E. Koshelev and A. A. Varlamov, Supercond. Sci. Technol. \textbf{27}, 124001 (2014).

\bibitem{Prange}R. E. Prange, Phys. Rev. B \textbf{1}, 2349 (1970).

\bibitem{Nakajima15}M. Nakajima, S. Uchida, K. Kihou, C. Lee, A. Iyo, and H. Eisaki, J. Phys. Soc. Jpn. \textbf{81}, 104710 (2012).

\bibitem{Hu15}D. Hu, X. Lu, W. Zhang, H. Luo, S. Li, P. Wang,G. Chen, F. Han, S. R. Banjara, A. Sapkota, A. Kreyssig, A. I. Goldman, Z. Yamani, C. Niedermayer, M. Skoulatos, R. Georgii, T. Keller, P. Wang, W. Yu, and P. Dai, Phys. Rev. Lett. \textbf{114}, 157002 (2015).

\bibitem{Swee}S. K. Goh, Y. Nakai, K. Ishida, L. E. Klintberg, Y. Ihara, S. Kasahara, T. Shibauchi, Y. Matsuda, and T. Terashima, Phys. Rev. B \textbf{82}, 094502 (2010).

\bibitem{Ishikado}M. Ishikado, K. Kodama, R. Kajimoto, M. Nakamura, Y. Inamura, S. Wakimoto, A. Iyo, H. Eisaki, M. Arai, and S. Shamoto, Physica C \textbf{471}, 643 (2011).

\bibitem{Babaev1}M. Silaev and E. Babaev, Phys. Rev. B \textbf{84}, 094515 (2011).

\bibitem{Babaev2}M. Silaev and E. Babaev, Phys. Rev. B \textbf{85}, 134514 (2012).

\bibitem{soto04}F. Soto, C. Carballeira, J. Mosqueira, M. V. Ramallo, M. Ruibal, J. A. Veira, and F. Vidal, Phys. Rev. B \textbf{70}, 060501 (2004).

\bibitem{carballeira3D}C. Carballeira, J. Mosqueira, M. V. Ramallo, J. A. Veira, and F. Vidal, J. Phys.: Condens. Matter \textbf{13}, 9271 (2001).

\bibitem{Klemm80}R. A. Klemm and J. R. Clem, Phys. Rev. B \textbf{21}, 1868 (1980).

\bibitem{Blatter92}G. Blatter, V. B. Geshkenbein, and A. I. Larkin, Phys. Rev. Lett. \textbf{68}, 875 (1992).

\bibitem{Hao92}Z. Hao and J. R. Clem, Phys. Rev. B \textbf{46}, 5853 (1992).

\bibitem{FVidal}F. Vidal, C. Carballeira, S. R. Curr\'as, J. Mosqueira, M. V. Ramallo, J. A. Veira, and {J. Vi\~na}, Europhys. Lett. \textbf{59}, 754 (2002).

\bibitem{ikeda89}R. Ikeda, T. Ohmi, and T. Tsuneto, J. Phys. Soc. Jpn. \textbf{58}, 1377 (1989).

\bibitem{ikeda90}R. Ikeda, T. Ohmi, and T. Tsuneto, J. Phys. Soc. Jpn. \textbf{59}, 1397 (1990).

\bibitem{kim92}D. H. Kim, K. E. Gray, and M. D. Trochet, Phys. Rev. B \textbf{45}, 10801 (1992).

\bibitem{MgB2}J. Mosqueira, M. V. Ramallo, S. R. Curr\'as, C. Torr\'on, and F. Vidal, Phys. Rev. B \textbf{65}, 174522 (2002).

\bibitem{NbSe2}F. Soto, H. Berger, L. Cabo, C. Carballeira, J. Mosqueira, D. Pavuna, and F. Vidal, Phys. Rev. B \textbf{75}, 094509 (2007).

\bibitem{MosqueiraEPL}J. Mosqueira, C. Carballeira, M. V. Ramallo, C. Torr\'on, J. A. Veira, and F. Vidal, Europhys. Lett. \textbf{53},  632 (2001).

\bibitem{Tl2223}J. Mosqueira, L. Cabo, and F. Vidal, Phys. Rev. B \textbf{76}, 064521 (2007).

\bibitem{TlPb1212}J. Mosqueira, and F. Vidal, Phys. Rev. B \textbf{77}, 052507 (2008).

\bibitem{intrinsic}J. Mosqueira, L. Cabo, and F. Vidal, Phys. Rev. B \textbf{80}, 214527 (2009).

\bibitem{Zaanen}J. Zaanen, Phys. Rev. B \textbf{80}, 212502 (2009).

\bibitem{Ullah1}S. Ullah, and A. T. Dorsey, Phys. Rev. Lett. \textbf{65}, 2066 (1990).

\bibitem{Ullah2}S. Ullah, and A. T. Dorsey, Phys. Rev. B \textbf{44}, 262 (1991).

\bibitem{Chong}S. Chong, S. Hashimoto, and K. Kadowaki, Solid State Commun. \textbf{150}, 1178 (2010).


\bibitem{Putzke}C. Putzke, P. Walmsley, J. D. Fletcher, L. Malone, D. Vignolles, C. Proust, S. Badoux, P. See, H. E. Beere, D. A. Ritchie, S. Kasahara, Y. Mizukami, T. Shibauchi, Y. Matsuda, and A. Carrington, Nat. Commun. \textbf{5}, 5679 (2014).

\bibitem{Chaparro} C. Chaparro, L. Fang, H. Claus, A. Rydh, G. W. Crabtree, V. Stanev, W. K. Kwok, and U. Welp, Phys. Rev. B \textbf{85}, 184525 (2012).

\bibitem{15Diaoarxiv} Z. Diao, D. Campanini, L. Fang, W. K. Kwok, U. Welp, and A. Rydh, arXiv:1503.04088.






\bibitem{fanfarillo}L. Fanfarillo, L. Benfatto, S. Caprara, C. Castellani, and M. Grilli, Phys. Rev. B  \textbf{79}, 172508 (2009).

\bibitem{orlova}N. V. Orlova, A. A. Shanenko, M. V.  Milo\ifmmode \check{s}\else \v{s}\fi{}evi\ifmmode \acute{c}\else \'{c}\fi{}, F. M.  Peeters, A. V. Vagov, and V. M. Axt, Phys. Rev. B \textbf{87}, 134510 (2013).

\bibitem{marciani}M. Marciani, L. Fanfarillo, C. Castellani, and L. Benfatto, Phys. Rev. B \textbf{88}, 214508 (2013).

\bibitem{klein}T. Klein, D. Braithwaite, A. Demuer, W. Knafo, G. Lapertot, C. Marcenat, P. Rodi\`ere, I. Sheikin, P. Strobel, A. Sulpice, and P. Toulemonde, Phys. Rev. B \textbf{82}, 184506 (2010).

\bibitem{serafin}A. Serafin, A. I. Coldea, A. Y. Ganin, M. J. Rosseinsky, K. Prassides, D. Vignolles, and A. Carrington, Phys. Rev. B \textbf{82}, 104514 (2010).

\bibitem{braithwaite}D. Braithwaite, G. Lapertot, W. Knafo, and I. Sheikin, J. Phys. Soc. Jpn. \textbf{79}, 053703 (2010).

\bibitem{LSCOanomaly} A. Lascialfari, A. Rigamonti, L. Romano', A. A. Varlamov, and I. Zucca, Phys. Rev. B \textbf{68}, 100505 (2003).

\bibitem{Lucia06}L. Cabo, F. Soto, M. Ruibal, J. Mosqueira, and F. Vidal, Phys. Rev. B \textbf{73}, 184520 (2006).

\bibitem{11Mosqueira} J. Mosqueira, J. D. Dancausa, and F. Vidal, Phys. Rev. B \textbf{84}, 174518 (2011).

\bibitem{YCaBaCuO}A. Lascialfari, A. Rigamonti, L. Romano', P. Tedesco, A. Varlamov, and D. Embriaco, Phys. Rev. B \textbf{65}, 144523 (2002).

\bibitem{Rigamonti}A. Rigamonti, A. Lascialfari, L. Roman\`o, A. Varlamov, and I. Zucca, J. Supercond. \textbf{18}, 763 (2005).

\bibitem{Rutgers}By combining the Rutgers expression $\Delta c/T_c=(\mu_0/2\kappa^2)(\partial H_{c2}^\perp/\partial T)^2_{T_c}$ with $\Delta c=2.8$~J/molK,\cite{Zaanen} and $\mu_0(\partial H_{c2}^\perp/\partial T)_{T_c}=-1.8$~T/K (Table~\ref{tableparameters}), one obtains $\kappa\approx30$. Then, by using $H_{c1}^\perp=(\ln\kappa/2\kappa^2)H_{c2}^\perp$ one finds $\mu_0(\partial H_{c1}^\perp/\partial T)_{T_c}\approx-3.4$~mT/K.

\bibitem{Bernardi}E. Bernardi, A. Lascialfari, A. Rigamonti, L. Roman\`o, M. Scavini, and C. Oliva, Phys. Rev. B \textbf{81}, 064502 (2010).



\end{references}

\end{document}